\documentclass[11pt, a4paper]{article}
\usepackage{jheppub}
\usepackage{amsmath,amssymb,amsfonts}
\usepackage{longtable}

\def\mylist{\begin{list}{}{\setlength{\leftmargin}{0.5in}
               \setlength{\listparindent}{-0.5in}
               \setlength{\itemindent}{\listparindent}}}

\newcommand{\bra}[1]{\langle#1\,|}
\newcommand{\ket}[1]{|#1\,\rangle}

\newcommand{\nD}[1]{\not D}

\title{Quantum Field Theory, Causal structures and Weyl transformations.}

\author{Denis Bashkirov}

\affiliation{Institut de Physique Th\'eorique Philippe Meyer\\
Laboratoire de Physique Th\'eorique, Ecole Normale Sup\'erieure\\
PSL Research University,\\
24 rue Lhomond, 75231 Paris Cedex 05, France\footnote{Current address} $\&$\\
Perimeter Institute for Theoretical Physics,\\
Waterloo, Ontario, ON N2L 2Y5, Canada\\}

\emailAdd{denis.bashkirov@lpt.ens.fr}

\abstract{ We suggest that in the proper definition, Quantum Field Theories are quantum mechanical system which 'live' on the space of causal structures ${\cal C}$ of spacetime. That is, for any QFT a Hilbert space ${\cal H}$ on which local operators live is assigned not for each Lorentzian metric $g$, but for each causal structure ${\cal C}$. In practice one uses 'conformal frames' which all provide equivalent descriptions of the same QFT. To put it differently, Quantum Field Theories only know about causal structure of spacetime, and not its full Lorentzian metric. The Weyl group and the local RG flow naturally arise when one compares equivalent descriptions in different conformal frames. This is reduced to the usual RG flow of coupling constants when one only compares descriptions in conformal frames related by spacetime-independent Weyl rescalings. We point out that in this picture minimal coupling of a QFT to the metric is inconsistent and comment on the necessary violation of the equivalence principle in the presence of scalars.}

\begin{document}

\maketitle

\flushbottom

\section{Physical motivation}

Let us start with the instrumentalist point of view. Suppose one is given a theory, that is a collection of local fields and the form of the action, for example, but is not told on what Lorentzian background one is placed. To what extent can one reconstruct the metric?

In order to more easily illustrate several important points let us be concrete and consider $\phi^4$ theory, although every statement is valid in general, without reference to any Lagrangians. 

The first step is obvious -- to determine the causal structure, one measures commutators of local observables at separate spacetime points.

To find the full answer one can measure correlation functions with separate-points insertions of local operators. However, not all information contained in them is 'physical': for any $c-$function $f(x)$ which is nowhere zero, operators ${\cal O}'(x)=f(x){\cal O}(x)$ are neither worse nor better than operators ${\cal O}(x)$ -- these are also local operators. This introduces ambiguity in correlation functions of local operators as devices to measure the Lorentzian metric since it is not clear a priori which of the two local operators corresponds to the local field $\phi(x)$ in the action for the $\phi^4$ theory. 

Indeed, if one uses a set of local operators $\{{\cal O}_i(x)\}$ to conclude/calculate that the metric is $g_{\mu\nu}$, then an alternative set of operators $\{{\cal O}'_a(x)\}$, local with respect to the first, meaning that
\begin{align}
\widehat{\cal O}'_a(x)=\sum_{i}f^i_a(x)\widehat{\cal O}_i(x)+f_i(x)\widehat 1
\end{align}
may find a different Lorentzian metric $g_{\mu\nu}'$. This is because it is not clear beforehand which of the two operators one should identify with the field $\phi(x)$ in the action if, f.e., one considers a $\phi^4$ theory. So the full metric may be ambiguous since it is not directly measurable.

 As we mentioned in the beginning, there is a part of the metric on which all observers, wether they prefer $\{{\cal O}(x)\}$ or $\{{\cal O}'(x)\}$, would agree: the causal structure: $g_{\mu\nu}$ and $g'_{\mu\nu}$ should produce the same causal structure on the Lorentzian manifold. Indeed, this is a question of wether commutators of local operators at separate points of spacetime are zero or nonzero -- this does not change with local fields redefinitions. All these commutators sit in correlations functions and form the invariant under change of descriptions, physical part, of them. For example, the mean value of the commutator of two local operators in a given state is the imaginary part of the time-ordered two-point correlator:
\begin{align}
\bra{\Psi}[{\cal O}(x),\tilde{\cal O}(y)]\ket{\Psi}=i\hbox{sign}(x^0-y^0)\hbox{Im}\bra{\Psi}T{\cal O}(x)\tilde{\cal O}(y)\ket{\Psi}
\end{align}

So, the metrics $g_{\mu\nu}$ and $g'_{\mu\nu}$ reconstructed from correlators of local operators ${\cal O}_i(x)$ and ${\cal O}_a(x)$ related by local transformations have necessarily the same causal structure. But that means that they are conformally related:
\begin{align}
g'(x)=\Omega^2(x)g(x).
\end{align}

Now, on one hand, this is obviously true in Conformal Field Theories. On the other hand, the physical arguments above do not seem to single them out from the class of all Quantum Field Theories. Indeed, in this operational setting it should be true that one should only be able to unambiguously reconstruct the metric only up to a conformal factor. We have seen that there is ambiguity in matching of local operators with local fields which allow the conformal equivalence to exist, but in general it may not be enough. To allow the conformal equivalence exist more generally, one needs to have more ambiguous information specializing a QFT to have at ones disposal -- these are parameters or coupling constants which specify a theory. We already know form the conventional RG  that those are ambiguous -- they 'run'. So, it must be them that change from one 'conformal frame' frame to another to allow for equivalent descriptions. In other words, to allow the metric to be unambiguous only up to a conformal factor. To put still differently, to guarantee that a QFT is defined not for the set of all Lorentzian metrics but for the set of all causal structures.

Of course, this point of view requires that all coupling constants be promoted to functions on spacetime. Not all choices of these coupling functions are allowed, though. For example, in flat Minkowski space, one should not allow the mass function $m^2(x)$ to be negative in some regions of space-time, as this will lead to super-luminal propagation in those region, that is, tho the loss of causality.

Thus we are lead to the suggestion that a QFT lives not on Lorentzian manifolds $\{{\cal M}, g\}$, but on the equivalence classes of Lorentzian manifolds $\{{\cal M}, {\cal C}\}$ with the same causal structure ${\cal C}$, or, equivalently on the equivalence classes $\{{\cal M}, g/Weyl\}$ whose representatives are related by Weyl transformations.

In practice, one picks a representative (a 'conformal' frame) with metric $g$ and coupling constants/parameters (now functions on space-time) $\lambda(x)$ and studies correlators of local operators ${\cal O}(x)$. One can equally choose another representative $g'=\Omega^2g$ with a different set of parameters $\lambda'(x)$ and study correlators with this metric and this set of coupling functions. The equivalence of descriptions is in the fact that there will be a choice of local operators ${\cal O}^{\Omega}_a(x)$ in the second frame related in a local way to the operators in the first frame ${\cal O}^{\Omega}_a(x)=\sum_{i}f^i_a(x){\cal O}_i(x)+f_a(x)1$ and $\lambda'(x)=f^{\Omega}(\lambda(x))$ such that triples $(g(x),{\cal O}(x),\lambda(x))$ and $(g'=\Omega^2g,{\cal O}^{\Omega}(x),f^{\Omega}(\lambda(x)))$ will yield the same correlators.

The case when there is a mixing with the identity operator in the relation
\begin{align}
{\cal O}^{\Omega}_a(x)=\sum_{i}f^i_a(x){\cal O}_i(x)+f_a(x)1
\end{align}
deserves a special mention -- this is the case of a conformal anomaly which is discussed in the next section.

Let us motivate our suggestion from a complimentary perspective by considering the standard Callan-Symanzik equation.\footnote{The form of the Callan-Symanzik equation we consider here is not quite standard: usually one keeps the expression for the metric the same and varies the dummy mass parameter $\mu$, while we keep $\mu$ fixed and vary the metric. These two points of view are equivalent -- see section 4.}

What is the meaning of the standard Callan-Symanzik equation for correlation functions of ${\cal O}$ in position space in a general QFT?
Suppose for simplicity that we consider a local operator which does not have an anomalous dimension. Then the statement of the Callan-Symanzik equation is the following:
\begin{align}
<\phi(x)\phi(y)>_{g,c}=\Omega^{-\Delta}\Omega^{-\Delta}<\phi(x)\phi(y)>_{\Omega^2g,c'(c,\Omega)}=F(x,y)
\end{align}
where $c'(c, \Omega)$ is a new value of the coupling constant $c$ obtained by running from the previous value by the RG under the rescaling by $\Omega$.

Now try to relate this to an experiment.  The correspondence between measuring devices and local fields (which is a computational tool), that is the correspondence between experiment and theory is not given a priori. One needs to match the two, so it is in general, for any QFT, is matching of pairs $(\hbox{parameters},{\cal O})$ (in a fixed state) to pairs $(\hbox{measuring apparatus},\hbox{outcomes})$. In our particular case it is the triple $(g,c,{\cal O})$ that needs to be matched to the measuring apparatus and measurements outcomes.

So suppose one measures correlation functions with a measuring device to be $F(x,y)$. Then one theorist says: this matches to measuring the field $\phi$, and so the metric and the coupling constant are $(g,c)$, but another theorist says: no, it corresponds to the field $\phi'=\phi/\Omega^\Delta$, and the metric is $g'=\Omega^2g$, and the coupling constant is $c'$ -- those are two interpretations on equal footing, and both interpretations are valid.

It is better to recast the (solution of) CS equation into its usual form:

\begin{align}
Z(\lambda)^{i_1}_{j_1}...Z(\lambda)^{i_k}_{j_k}<\phi^{j_1}(x_1)...\phi^{j_k}(x_k)>_{\lambda^2g,c_a(\lambda)}=const(\lambda)=
<\phi^{i_1}_0(x_1)...\phi^{i_k}_0(x_k)>
\end{align}

Here $\{c_a\}$ is the set of coupling constants, and the RHS is interpreted as mysterious 'bare' operators. They are actual expectation values of local operators ${\cal O}(x)$ which live in the Hilbert space ${\cal H}_{\cal C}$ and correspond to the concrete measuring device, while $\phi^i(x)$ are local fields which come with each conformal frame, so they part of a frame dependent description. We measure the 'bare' correlators -- expectation values of physical local operators ${\cal O}(x)$ corresponding to local measuring apparatus in some state $\ket{\Psi}_{\cal C}$. They are obviously frame-independent.

So
\begin{align}
<\phi^{i_1}_0(x_1)...\phi^{i_k}_0(x_k)>_{\Psi}=\bra{\Psi}T\widehat{\cal O}^{i_1}(x_1)...\widehat{\cal O}^{i_k}_0(x_k)\ket{\Psi}
\end{align}

Even if in some conformal frame all $Z_i^j$ become the identity matrix, it doesn't mean the metric corresponding to this frame is the 'real' metric -- it's again a matter of convention. One physicist might like the identity matrix, another -- some other matrix. The point is the very fact that the expectation value CAN be interpreted differently, and the matter of easiness/convenience is irrelevant in principle (but not for practical calculation, of course).

The conclusion is that the standard RG invariance implies that metric itself is not physical -- the physical part is only the scaling equivalence class.

In a CFT the redundancy is enlarged to the full Weyl group, which means that physical (unambiguous) is only the causal structure.

The redundancy cannot be larger than the Weyl group because all matchings agree on wether commutators are zero or not.

But is the scaling redundancy the full redundancy in non-conformal QFTs?

\begin{itemize}
\item First of all, unlike the CFT case, there is no clear physical expalnation/motivation for this redundancy.
\item In the spirit of locality, it seems natural that the redundancy should work locally.
\end{itemize}

Let us return to the first example of two-point correlator with a single coupling $\lambda$ for simplicity. If it works locally it means that the triples $(g(x),c(x),\phi(x))$ and $(\Omega^2(x)g(x),\lambda^\Omega(x),\phi^\Omega(x))$ should give the same correlator $F(x,y)=\bra{\Psi}T{\cal O}(x){\cal O}(y)\ket{\Psi}$. Here $c^\Omega(x)$ is a local function $c^\Omega(x)=f(c(x),\Omega(x))$ (and their derivatives) and $\phi^\Omega(x)$ is a local rescaling of $\phi(x)$.

But then it is nothing else then the statement of the existence of a local RG flow!

Furthermore, now this local redundancy has a physical explanation/motivation: the only unambiguous information that any QFT contains about Lorentzian metric is its causal structure. The rest is ambiguous -- this is how the local RG flow appears -- it's just comparison of equivalent descriptions in different 'conformal' frames.

It is the correlators of local operators that are frame independent (for any tuple of spacetime points $(x_1,...,x_k)$ it is just a number when $\mu$ is fixed) and what all frames $(g(x),c(x),\phi(x))$ give.

\section*{Coupling functions/sources}

Coupling functions play a double role in QFT. On one hand, a set of coupling functions ${\cal J}=\{J_I(x)\}$ in a fixed conformal frame defines a particular theory for a fixed causal structure. On the other hand they serve as sources for correlation functions of local operators through the relation
\begin{align}
\bra{\Psi}T{\cal O}_1(x_1){\cal O}_2(x_2)...{\cal O}_k(x_k){\cal O}\ket{\Psi}_{\{g,J\}}=\frac{\delta^k}{\delta J_1(x_1)\delta J_2(x_2)...\delta J_k(x_k)}\bra{\Psi}{\cal O}\ket{\Psi}_{\{g,J\}}
\end{align}

Here ${\cal O}$ stands for some insertions of some other operators. In particular, an insertion of the stress-tensor corresponds to a variation with respect to the metric. This is why a QFT should be defined on general (or at least generic) Lorentzian manifolds even if at the end of the day one only studies it in Minkowski space. We will have more to say about this in Section 4.

A predictive QFT is characterized by a finite set of coupling functions (modulo Weyl equivalence), while every QFT contains infinitely many local operators to each of which there corresponds a source. The Weyl group acts on this infinite set of sources, so a predictive QFT is characterized by the property that there is  an 'almost fixed point' in this infinite-dimensional space -- a 'fixed subspace' of a finite dimension.

Furthermore, if there is a true fixed point in this space, then it is natural to make it the origin for the sources, that is, to introduce the convention that all the coupling functions are zero at this point which corresponds to a Conformal Field Theory.

Such Conformal Field Theories exist in odd number of dimensions, but strictly speaking there are no such fixed points in even dimension due to the conformal anomaly. More precisely, in odd dimensions there is a conformal anomaly due to spacetime-dependent coupling functions but it can be set to zero for a certain choice of coupling functions (zero coupling functions by the above convention). In even dimensions this cannot be done. So one can have at best an Anomalous Conformal Field Theory.

A conformal anomaly is the situation when there is an additive (spacetime-dependent) shift in the coupling function for the identity operator (it can be called 'vacuum energy density'). Due to this it is not possible to fix this coupling function for the entire conformal class, so there is no fixed point, but there is a pretty harmless fixed 'line' which does not affect correlators at separate points. Due to the double role of coupling functions this is exactly related to the presence of mixing with identity operator discussed in the previous section.

It should be stressed that above we considered arbitrary Quantum Field Theories, and not just Conformal Field Theories.

Now, states live on causal structures/conformal classes, one just probes/describes them differently (but equivalently) in different conformal frames. So if one considers states in the Schrodinger picture as functionals
\begin{align}
\Psi_{\{g,J\}}[\phi(x)],
\end{align}
one has to make sure this functional is not changed under conformal transformation of sources $\{g,J\}$ and fields $\phi(x)$.

\section{Units and dimensional analysis in Quantum Field Theory}

In Quantum Field Theory there is only one unit -- the lengths unit, but more often one talks about the mass unit which is just the inverse of length. All operators have definite length (or, equivalently, mass) dimensions: every operator is measured in a power of length which is called its length dimension equal to minus mass dimension.

There are two ways to implement dimensional analysis. The first is what one finds in classical physics and in quantum mechanics (of finitely many degrees of freedom), when for a dimensionless quantity $F$ which is a function of dimensionful variables $\{R_i\}$ there is a single formula
\begin{align}
F=F(R_i).
\end{align}
Consider for concreteness the case of two variables $R_1$ and $R_2$ with length dimension $1$ and $2$ respectively. Then by dimensional analysis the dimensionless quantity $F$ is a function of the dimensionless ratio $R_1^2/R_2$:
\begin{align}
F=F(R_1,R_2)=f(R_1^2/R_2).
\end{align}

This is not the only option to make the dimensional analysis work. The second possibility is that for a dimensionless quantity $F$ which is a function of dimensionful variables $\{R_i\}$ there are infinitely many expressions -- one for each choice of units. Let us denote these functions by $F_\xi(R_i)$ - there is one parameter family of functions with $\xi$ running over the real line ${\mathbb R}$.

Those are expressions for numerical values of all quantities: one for when one measures in meters, one for centimeters, one for millimeters etc. Each formula describes the relation between numerical values of all quantities. Because in each 'unit frame' (choice of length unit) all variables are just dimensionless variables, each function $F_\xi(R_i)$ may be more complicated than $F(R_i)$ from the previous option, but since they all express a dimensionless quantity which is independent of choice of units, there must be a relation between $F_\xi(R_i)$ for different $\xi$ which ensures this.

Return to the previous example of two dimensionful variables $R_1$ and $R_2$ with length dimension $1$ and $2$ correspondingly.

Suppose we measure everything in centimeters and find the following relation between $F,R_1,R_2$ in these units:
\begin{align}
F=10R_1R_2.
\end{align}
Next we switch to using millimeters. The numerical values for $R_1$ get multiplied by 10 and for $R_2$ -- by 100 while numbers for $F$ stay the same. This means that in millimeters unit the formula has the form
\begin{align}
F=10\times10\times 10^2R_1R_2.
\end{align}
The point is that for a chosen unit the formula may take any form, but then for all other choices of units (in all other 'unit frames')it is fixed.

So, in this example
\begin{align}
F=\xi R_1R_2
\end{align}

where under change in units $R_1\to\lambda R_1$, $R_2\to\lambda^2 R_2$, $F\to F$, so that $\xi\to\lambda^{-3}\xi$. The parameter $\xi$ transforms as the third power of a mass parameter, so it is convenient to denote it $\mu^3$. Then the formula takes form
\begin{align}
F=F(R_1,R_2;\mu)=\mu^3R_1R_2
\end{align}
Now if one treats the parameter $\mu$ as a dummy mass, one is back to the first option of implementing dimensional analysis at the expense of introducing a dummy mass parameter $\mu$.

This second option is more general than the first one, obviously, as it allows for more functional dependence on physical variables $\{R_i\}$. Furthermore, it contains the first option since the original formula is reproduced if one requires the dependence on $\mu$ to be trivial.

More importantly, it is the second option that is realized in QFT: the introduction of the dummy mass parameter $\mu$ is forced by the renormalization procedure. A finite $\mu$ remains at the end of the day after the theory is made finite -- this is the essence of the renormalization procedure.

\section{Inconsistency of minimal coupling to geometrical background}

As was noted above a QFT must be defined on a general Lorentzian manifold in order for variations with respect to the metric produce a stress-tensor.

Naively, it seems that given a QFT in flat space there are many ways to put it on curved backgrounds. To be concrete, consider the massive $\phi^4$ scalar theory from \cite{CCJ} with flat space action
\begin{align}
S=\int d^4x(\eta^{\mu\nu}\partial_\mu\phi(x)\partial_\nu\phi(x)+m^2\phi^2(x)+\lambda\phi^4(x)).
\end{align}
It seems that there is a preferred way to put this theory on a curved spacetime -- the minimal coupling. One just promotes the Minkowski metric to a general metric and gets
\begin{align}
S=\int d^4x\sqrt{g}(g^{\mu\nu}\partial_\mu\phi(x)\partial_\nu\phi(x)+m^2(x)\phi^2(x)+\lambda(x)\phi^4(x)).
\end{align}

According to the suggestion that a QFT 'lives' on causal structures rather than metrics, this does not correspond to putting a single $\phi^4$ theory on arbitrary curved background: the action does not preserve its form under conformal transformations because of the kinetic term and therefore is not defined on causal structures only. Thus it is putting a family of $\phi^4$ theories (parameterized by the metric $g_{\mu\nu}$ up to a constant scale factor) each in a single curved metric (up to a constant scale factor), and not a single theory in arbitrary background.

Thus when one performs a variation with respect to the metric to get an insertion of the stress-tensor, one does not vary the metric for a fixed theory, but performs a variation along a section in the space of $\phi^4$ theories and metrics. Correspondingly, one expects that this stress-tensor is not well-defined even in Minkowski space. Furthermore, one expects that problems come from the trace part of the stress-tensor, as it is this part that corresponds to varying the conformal factor of the metric, and it is the relation between conformal frames which does not work for this minimal coupling.

This is exactly the result of \cite{CCJ} who found that such tensor does not exist because it has infinite matrix elements due to the trace part.

The correct way to put this theory on curved backgrounds is to include the conformal coupling to the curvature
\begin{align}
S=\int d^4x\sqrt{g}(g^{\mu\nu}\partial_\mu\phi(x)\partial_\nu\phi(x)+\frac{1}{6}R(x)\phi^2(x)+m^2(x)\phi^2(x)+\lambda(x)\phi^4(x))
\end{align}
In this case, one indeed varies metrics for a fixed $\phi^4$ theory. The same authors \cite{CCJ} showed that the stress-tensor obtained by varying this action (called new improved stress-tensor) has finite matrix elements. Because this direct coupling to curvature does not disappear in local inertial frames, the equivalence principle is violated in our interpretation of QFT in the presence of scalars (like in the Standard Model).

A quick remark about scale but not (anomalous) conformal theories is appropriate at this point. Consider the example of the free ${\mathbb R}$ (not $U(1)$) Maxwell theory in three dimensions. It is scale but not conformally invariant unless one introduces monopole operators \cite{ERN}. On a topologically trivial Lorentzian manifold one can dualize this theory to the theory of free massless scalar. Not including monopole operators allows only minimal coupling of this scalar to background metric which is not consistent. Presumably, on the flat spacetime this will show in some correlation functions with stress-tensor insertions.

It is possible that the same problem happens for other cases of scale but not conformal theories.

\section{Callan-Symanzik equation}
In this section we show that for conformal frames related by spacetime-independent conformal factors the equivalence of descriptions is equivalent to the requirement that separate-points correlators satisfy Callan-Symanzik equation.

In fact, this was done 25 years ago by Hugh Osborn \cite{Osb}\footnote{For a recent summary and some extension see \cite{Retal}}. Namely, instead of correlators he considered the generating functional $W[g,J;\mu]$ for the connected correlators. In the presence of massless degrees of freedom this functional is not defined for any background metric and sources -- one needs to choose them so that ground state degeneracy is lifted, then take variational derivatives to obtain connected correlators (which always exist) and only after that take the limit of sources/metric to land in the desired vacuum. In this description the double role of coupling functions is obvious.

For completeness, here we give a quick recap of how it goes.

The requirement of equivalence of descriptions in different conformal frames is the requirement of invariance of $W[g,J;\mu]$ under conformal transformations up to a conformal anomaly:
\begin{align}
W[\Omega^2(x)g(x),J^\Omega(x);\mu]=W[g(x),J(x);\mu]+{\cal A}[g(x),J(x);\Omega(x)],
\end{align}
where the conformal anomaly ${\cal A}[g(x),J(x);\Omega(x);\mu]$ is a local functional of its arguments. The dummy mass parameter is not a function on spacetime!

If one introduces the source $\Lambda(x)$ (vacuum energy density) for the identity operator and defines $W[g(x),J(x),\Lambda(x);\mu]\equiv W[g(x),J(x);\mu]+\int\Lambda(x)$, the equation can be recast into a manifestly form-preserving form.
\begin{align}
W[\Omega^2(x)g(x),J^\Omega(x),\Lambda^\Omega(x);\mu]=W[g(x),J(x),\Lambda(x);\mu],
\end{align}
with $\Lambda^\Omega(x)=\Lambda(x)-a(x)$ where $a(x)$ is the density of the local anomaly functional ${\cal A}[g(x),J(x);\Omega(x)]$.

We will use the first option with the explicit anomaly.

The second equation expresses the requirement of dimensional analysis: since $W$ is dimensionless, it should not depend on the choice of units. Because of diffeomorphism covariance the change of units affects the arguments of $W$ in the following way:
\begin{align}
\mu\to\lambda\mu,\qquad g(x)\to\lambda^{-2}g(x),\qquad J(x)\to\lambda^{m[J]}J(x).
\end{align}
Here $m[J]$ is the mass (inverse scale) dimension of $J$. If at small scales the flow ends in a CFT for which the coupling functions are by convention take to be zero, then mass dimensions of all local operators are just their UV conformal dimensions. Thus $m[J]=\Delta[J]$.

So, the second equation  which ensures that dimensional analysis works is
\begin{align}
W[\lambda^{-2}g(x),\lambda^{\Delta[J]}J(x),\lambda\mu]=W[g(x),J(x),\mu].
\end{align}
Next, for small values of $J(x)$ (in the units of $\mu$)(close to the UV CFT) the action of the Weyl group on the coupling functions are
\begin{align}
J^{\Omega(x)}(x)=\Omega^{\Delta[J]}(x)J(x)+\gamma[J(x),\Omega(x)]J,
\end{align}
where the anomalous scaling $\gamma[J,g]$ is of order $J$ (in the case when $\Delta[J]=0$ the first term is absent and the second does not have $J$ multiplying $\gamma$). This corresponds to the anomalous dimension of the corresponding local operator. There is in general a mixing of operators which we omit in order not to clutter the notation.

One should stress that the scaling of the metric in the first equation is physical, while in the second equation it is just a change of length unit.

Next, one considers a 'diagonal' section of this physical and dummy scalings so that the explicit expression for the metric stays the same, combines the two equation and arrives at the Callan-Symanzik equation for scale transformations:
\begin{align}
(\mu\frac{\partial}{\partial\mu}+\int d^4x\gamma[J(x),\lambda]J\frac{\delta}{\delta J(x)})W[g,J;\mu]={\cal A}[g,J;\lambda]
\end{align}
Taking variational derivatives at separate spacetime points one gets the  Callan-Symanzik equation for connected correlation functions. The anomaly contributes only to the 1-point insertion in a CFT, but for any QFT it is zero (for constant rescalings) in flat spacetime.

We should stress that although the metric does not appear to be varied in this equation, it actually is just an appearance coming from combining physical variation with a change of length unit.

\section{Summary}

In this paper we suggest the following definition of a Quantum Field Theory:
\begin{itemize}
\item A QFT is a quantum mechanical system for which for every causal structure ${\cal C}$ of spacetime there correspond a Hilbert space ${\cal H}_{\cal C}$ together with local fields ${\{\cal O}_i\}$ (operators when smeared over a finite open set) acting on it. Elements of the Hilbert space ${\cal H}_{\cal C}$ are states $\ket{\Psi}_{\cal C}$.
\item The set of causal structures on spacetime has a description as the quotient of the set of all lorentzian metrics by the action of the Weyl group. Thus to describe a causal structure one can pick a representative in the equivalence class -- a Lorentzian metric and describe a QFT in this 'conformal' frame. All such descriptions are equivalent (though, of course, some may be easier).
\item In practice a QFT is described by the set of local (and also higher-dimension) fields together with a parameter  -- in each conformal frame this parameter is a set of coupling functions $J_A(x)$ on spacetime. All conformal frames are equivalent and describe the same physics. The change of a conformal frame induces the local action of the Weyl group on the space of coupling functions and local fields: the two description $(g(x), J(x), {\cal O}(x))$ and $(\Omega^2(x)g(x),J^{\Omega(x)}(x),{\cal O}^{\Omega(x)}(x))$ of the QFT corresponding to two conformal frames are equivalent -- they give the same correlation functions. This has the interpretation of a local RG flow.
\item The usual RG flow corresponds to comparing descriptions in conformal frames related by a constant conformal factor $\Omega$.
\item This must be supplemented by additional physical requirements like unitary time evolution in conformal frames which allow global time-like Killing vectors (in those frames it makes sense to take coupling functions to be time-independent), and the choice of coupling functions which do not violate causality (like locally negative mass function $m^2(x)$ in flat conformal frame.)
\end{itemize}

\section{Acknowledgements}

This research is supported by the
Perimeter Institute for Theoretical Physics. Research at the Perimeter Institute
is supported by the Government of Canada through Industry Canada
and by Province of Ontario through the Ministry of Economic Development
and Innovation.

We would like to thank the hospitality of
the Simons Center for Geometry and Physics where work that has lead to this publication started during the 2015
Simons Summer workshop in mathematics and physics.

\end{document}